\documentclass[11pt, epsf]{article}
\usepackage{epsfig}
\def\ga{\mathrel{\raise.3ex\hbox{$>$\kern-.75em\lower1ex\hbox{$\sim$}}}}
\def\la{\mathrel{\raise.3ex\hbox{$<$\kern-.75em\lower1ex\hbox{$\sim$}}}}

\def\I_M{{I_{\scriptscriptstyle M\times M}}}

\newcommand{\be}{\begin{equation}}
\newcommand{\ee}{\end{equation}}
\newcommand{\f}{\frac}
\newcommand{\bea}{\begin{eqnarray}}
\newcommand{\eea}{\end{eqnarray}}

\begin{document}

\thispagestyle{empty}
\rightline{IP/BBSR/2003-01}

\vskip 2cm \centerline{ \Large \bf Nonsingular Cosmologies from Branes}

\vskip .2cm

\vskip 1.2cm

\centerline{ \bf Anindya Biswas, Sudipta Mukherji and Shesansu
Sekhar Pal}
\vskip 10mm \centerline{ \it Institute of Physics, 
Bhubaneswar-751 005, India} 
\vskip 1.2cm

\centerline{\tt  anindyab, mukherji, shesansu@iopb.res.in,
}

\vskip 1.2cm

\begin{quote}

We analyse possible cosmological scenarios on a brane where the brane
acts as a dynamical boundary of various black holes with  anti-de 
Sitter or de Sitter asymptotic. In many cases, the brane is found to 
describe completely non-singular universe. In some cases, quantum gravity
era of 
the brane-universe can also be avoided by properly tuning bulk parameters.
We further discuss the creation of a brane-universe by studying its wave 
function. This is done by employing Wheeler-De Witt equation in the
mini superspace formalism. 
\end{quote}

\newpage
\setcounter{footnote}{0}
\tableofcontents

\section{Introduction}

Motivated by string theory, the AdS/CFT correspondence and the hierarchy 
problem in particle physics, brane-world models have been a focus of 
interest in recent years \cite{HW} - \cite{RSt}. In these models, our 
universe is realized as a 
boundary of a higher-dimensional space-time. In this context, a well 
studied example is when the bulk is an AdS space. The gravitational 
interaction among matter on this brane is found to be described by 
standard laws when one considers distance scale much larger than the AdS 
length scale \cite{gata}. 

In the cosmological context, many authors have considered (see for example 
\cite{TN} - \cite{cm}) 
the embedding of a four dimensional Friedmann-Robertson-Walker (FRW) 
universe in five dimensional bulk geometry. The bulk is often described by 
AdS or AdS 
Schwarzschild black hole metric. In the later case, the mass of the 
black hole is found to act effectively as an invisible energy density on
the brane with the same equation of state of radiation matter. This, 
however, is found to have a nice interpretation in terms of AdS/CFT 
correspondence. 
The correspondence allows one to reinterpret the AdS-Schwarzschild 
geometry as a source for a four dimensional CFT at finite temperature. 
As pointed out in \cite{Gub} - \cite{SV}, it is easy then to understand 
why the 
radiation-dominated FRW universe emerges. This is because all CFTs have 
the same equations of state up to numerical constants and hence the FRW 
equation takes the same form as in the case of radiation dominated 
universe. Replacing the bulk now by a five 
dimensional AdS-Reissner-Nordst\"orm black hole may seem like 
a straight forward generalization. However, as it was found in \cite{MP}, 
such 
geometry has far reaching consequences on the brane evolution. First, FRW 
equation 
now describes the brane as dominated by induced radiation and ``stiff'' 
matter. However, in the equation, the stiff matter 
contribution comes with an opposite sign\footnote{This was first noticed 
in \cite{MV1}.}. 
As a result, the universe (flat, 
open and closed) is never found to shrink to zero size. It is important 
to note that the bulk electric charge plays the role of a ``regulator'' on 
the brane. As we take the charge to zero, big bang singularities 
are found to reappear\footnote{Some earlier work on brane cosmology can 
be found in \cite{MP, BM}.}.

Here, in this paper, we further analyse the $D$ dimensional brane 
evolution 
in various $D+1$ dimensional AdS or dS black hole backgrounds of
various topologies. We first study these backgrounds carefully and find 
critical values of mass and other parameters  in order these geometries to
correspond to black holes with non-degenerate horizons. We believe
that some of these explicit results here are new. We then study the brane
dynamics in these geometries. It is found to be described by FRW equation
which has
contributions  from induced radiation matter, stiff matter and an
effective cosmological  constant. This constant depends on a certain
combination of bulk  cosmological constant and brane tension. We then
analyse the classical  dynamics of the brane by solving the FRW equation
in case of  
vanishing and nonvanishing effective cosmological constant. 
We find that all open, flat and closed universes have {\it no 
big bang} singularities when the bulk has a nonvanishing electric charge. 
For the closed universe, the size is 
bounded from above and from below. On the other hand, for open and flat 
universe, the brane starts with finite radius and expands for ever.
We also find  here that the {\it minimum} radius of the 
universe can be made sufficiently {\it large} by tuning the bulk mass 
and charge to large values. This, in turn, means that for such universes 
we can {\it avoid} reaching a quantum gravity era in the past. This scenario, 
in a more general set up, is analysed recently in \cite{EM}. We then
analyse the brane universe when there is an effective
(negative or positive) cosmological
constant on the brane. Though the FRW equations here can not be solved
analytically in a closed form, we study the equation by numerical
means. The non-singular nature of the solutions is also found to persist
in this case. However, is all the above cases, universe is found to be
singular when the charge is set to zero. We, therefore,
turn our attention to the fate of these singularities from the perspective 
of ``quantum cosmology''.  By employing the Wheeler De Witt (WDW) equation 
in the mini superspace formalism, we find that the minimum size of the 
brane universe is of the order of the Planck length. We feel that this may 
indicate that the universe stabilizes against collapse due to 
quantum effects. However, we may add here that quantum cosmology at
this present stage has certain inherent ambiguities. These are  associated
with the orderings of operators, choice of boundary conditions of the wave
functions etc. That is why our results here will be more of speculative in
nature.

This paper is organized as follows. In the next section, we critically
analyse various topological AdS and dS black hole backgrounds. Considering
brane as a dynamical boundary of these geometries,
we set up  brane equation of motion by studying the
brane effective action. In section 3, we analyses classical evolution of 
the brane by explicitly solving brane equation of motion. 
\footnote{Some cosmological scenarios on the brane have been reviewed
in \cite{NOO}.} We then critically analyse various physical properties of
our solutions with a particular emphasize on their early and late time
behavior. Section 4 of this paper deals with the brane 
dynamics from the WDW perspective. We solve relevant WDW equations , in 
mini superspace, for tensionless branes and analyse its behavior at short 
distance scales. For branes with tension, one needs to resort to some 
approximate methods. We study these branes in section 5 via WKB 
approximation. The paper ends with a discussion of our results.

\section{Brane effective action\label{EPS}}

\noindent{\bf Bulk with AdS asymptotics}

The dynamics of a $D$ dimensional brane inside a bulk gravitational field
is dictated by the following bulk-boundary action with a negative
cosmological constant $\Lambda=\frac{-D(D-1)}{2l^2}$
\begin{eqnarray}
S &=& {1\over{16 \pi G_{D+1}}}\int_{\cal{M}} d^{D+1} x {\sqrt g}\Big(R 
    -F^2+
     {D(D-1)\over{l^2}}\Big) \nonumber\\
    &-& {1\over{8 \pi G_{D+1}}}\int_{\partial{\cal M}}
      d^Dx {\sqrt\gamma}{\cal{K}}
    - {T\over{8 \pi G_{D+1}}}\int_{\partial{\cal M}} d^Dx{\sqrt{\gamma}},
\label{bulk-bdry}
\end{eqnarray}
where $F^2$ is the Maxwell kinetic energy term and ${\cal K}$ is the trace 
of the extrinsic curvature ${\cal
  K}_{\mu\nu}$ taken with respect to the 
{\it induced} metric on the brane $\gamma_{\mu\nu} = g_{\mu\nu} 
- n_\mu n_\nu$. This term is required for a well-defined variational
principle on the space time boundary \cite{GH}. Here $n_\mu$ is the unit 
normal vector to the brane. 
Furthermore, $T$ in (\ref{bulk-bdry}) is the brane tension and this
term is required in order to get a finite action and stress tensor
\cite{BK, KLS}\footnote{In general, one should add more derivative
  terms to (\ref{bulk-bdry}), see \cite{KLS}.}. Variation of (\ref{bulk-bdry})
with respect to the
induced metric gives us 
\begin{equation}
{\cal K}_{\mu\nu}=\f{T}{D-1} \gamma_{\mu\nu}.
\label{eom}
\end{equation}

A large class of solutions of bulk equations of motion can be represented
by 
\begin{equation}
ds^2 = - h(a) dt^2 + {da^2\over{h(a)}} + a^2 \tilde\gamma_{ij} dx^i dx^j,
\label{bulkmet}
\end{equation}
where 
\begin{equation}
h(a) = k + {a^2\over l^2} - {\omega_D M\over {a^{D-2}}} + {(D-1) \omega_D^2
      Q^2 \over{8 (D-2) a^{2 D -4}}}, ~~\omega_D = {16 \pi
      G_{D+1}\over{(D-1) V_{D-1}}}.
\label{valueh}
\end{equation}
Here, $k = 0, \mp 1,$  correspond to flat, hyperbolic and spherical
 geometries of
$D$ dimensional subspace for a given $a$. $\tilde\gamma_{ij}$ is the 
metric for 
a
constant curvature manifold $M^{D-1}$ with volume 
$V_{D-1} = \int d^{D-1} x \sqrt{\tilde \gamma}$. As can easily be seen 
from 
(\ref{bulkmet}) and
(\ref{valueh}), for $M = Q = 0$, the bulk is a
simple $D$ dimensional AdS space. For only  $Q = 0$, the bulk is an
AdS Schwarzschild black hole, while for all the parameters non-zero and 
within certain domain, the
background corresponds to charged AdS black holes. The parameters $M$ and
$Q$ can then be identified with Arnowitt-Deser-Misner mass and charge
respectively. Some analysis of the  causal structure of this metric
has been performed in \cite{CEJM,CS}. 

We would like to comment on the horizon structure of the metric 
(\ref{bulkmet}). First, note that this metric asymptotically is anti-de 
Sitter for all values of parameters. However, for the metric to describe 
the exterior of black hole with non-degenerate horizon, we need to 
restrict the parameters. To have such a geometry, we would like  
$a^{(2D -4)} h(a)$ to have a simple root at $a = a_0$ such that 
$h(a)$ is positive for $a >a_0$. This gives a lower bound on the 
parameter $M$. This can be seen as follows. Let $h(a)$ be zero for 
$a = a_0$. Therefore, we can invert the relation to get mass $M$ as a 
function of $a_0$ and other parameters as
\begin{equation}
M = {a_0^{(D -2)}\over \omega_D} \Big[k + {a_0^2\over l^2}
+ {(D-1) \omega_D^2 Q^2 a_0^{-2 D +4} \over{8(D-2)}}\Big].
\label{mroot}
\end{equation}
Now, by taking first and second derivative of (\ref{mroot})
with respect to $a_0$, we see that $M$ has a minimum $M^{\rm crit}$
at $a_0^{\rm crit}$ when
\begin{equation}
8 D {(a_0^{\rm crit})}^{2 (D-1)} + 8 (D-2) k l^2 {(a_0^{\rm crit})}^{2D 
-4}
- l^2 \omega_D^2 (D-1) Q^2 = 0.
\label{mmin}
\end{equation}
$M^{\rm crit}$ can therefore be determined by first solving $a_0^{\rm 
crit}$ from (\ref{mmin}) and then substituting it into (\ref{mroot}).
If we now define $a_{H} (M,Q)$ as the larger of the positive solutions, we 
will have the following inequality:
\begin{equation}
a_0^{\rm crit} (Q) \le a_H (M,Q) < \infty, ~{\rm as}
~M^{\rm crit}(Q) \le M < \infty.
\end{equation}
Though, in higher dimension, expressions of $M^{\rm crit}$ and
$a_0^{\rm crit}$ are either not illuminating or hard to obtain, but for 
$D=3$, we have
\begin{eqnarray}
M^{\rm crit} &=& {\sqrt{2}} (-k^2 l^2 + 3 \omega_3^2 Q^2 
+ kl {\sqrt{k^2 l^2 + 3\omega_3^2 Q^2}})\over{{\sqrt{27}}\omega_3 
{{\sqrt{l}} (-k l 
+ {\sqrt{k^2 l^2 + 3 \omega_3^2 Q^2}}})^{1\over 2} }, \nonumber\\
a_0^{\rm crit} &=& {\sqrt{-{k l^2\over 6} + {l^2\over 6} {\sqrt{k^2 + {3 
\omega_3^2 Q^2\over l^2}}}}}.
\label{fourd}
\end{eqnarray}
We note here that $a_0^{\rm crit} \rightarrow 0$ as $Q \rightarrow 0$. 
For $D=4$, that is when the bulk is five dimensional, it is also 
possible to find $M^{\rm crit}$ and $a_0^{\rm crit}$ explicitly. However, 
the expressions are very big except for $k=0$. In this case,
\begin{eqnarray}
M^{\rm crit} &=& {3^{2\over 3} \Big( 5  \omega_4 Q^2
+ {\sqrt{l^4 \omega_4^2 Q^4}}\Big)\over
{16  \Big(l^2 \omega_4^2 Q^2 + {\sqrt{l^4 \omega_4^4 
Q^4}}\Big)^{1\over 3}}},\nonumber\\
a_0^{\rm crit} &=& {3^{1\over 6}\over 2}\big( l^2 \omega_4^2 Q^2 +
{\sqrt{l^4 \omega_4^4 Q^4}}\big)^{1\over 6}.
\label{fbrane}
\end{eqnarray}
As we would like to find the dynamics of a $D$ dimensional brane moving in 
(\ref{bulkmet}), it is instructive to find out the {effective} action 
that controls the dynamics. For this purpose (and hence in the rest of the 
section), we do not need to explicitly specify the functional form of 
$h(a)$. As we will see, effective action describing the boundary dynamics 
can be expressed in terms of $h(a)$.

The extrinsic curvature, ${\cal K}_{\mu\nu}$, can be calculated following
\cite{MV}. Those are
\be
{\cal K}^{\theta_i}_{\theta_i}=\f{{\sqrt{h(a)+{\dot a}^2}}}{a},
\ee
and
\be
{\cal K}^{\tau}_{\tau}=\f{h^{'}(a)+2{\ddot a}}{2 {\sqrt{h(a)+{\dot a}^2}}},
\ee
where $h^{'}(a)=\f{dh(a)}{da}$ and $i=1,\ldots, D-1$, 
 $\theta_i$'s are the angular coordinates,  ${\dot a}=\f{da}{d\tau}$
 and $\tau$ denotes the proper time as measured along the brane
 world volume. Details of these computations can be found in many 
earlier literature, see for example \cite{MV}. Inserting these values of 
extrinsic curvature in (\ref{eom}), we get  
\be 
\f{\sqrt{h(a)+{\dot a^2}}}{a}=\f{T}{D-1}.\\
\label{eom1}
\ee 

The dynamics of the brane can be captured through an effective Lagrangian
$L$. Clearly, the only degree of freedom  in $L$ will be $a$ - so the
Lagrangian which we are about to write down is a construction in the
mini-superspace. It is given by
%
\begin{equation}
L = M_p\Bigg[\dot a \sinh^{-1} \Bigg({\dot a\over {\sqrt {h(a)}}}\Bigg) -
{\sqrt{h(a) +
\dot a^2}}
+ {T \over{D-1}}a\Bigg],
\label{effectivel}
\end{equation}
where $M_p$ is Planck mass, a dimension full constant.  The 
Euler-Lagrange 
equation for $a$ then is given by
\begin{equation}
\f{h^{'}(a)+2\ddot a}{2\sqrt{h(a)+{\dot a}^2}}=\f{T}{D-1}. 
\label{eqm}
\end{equation}
$L$ then reproduces the correct equation of motion of the brane
(\ref{eom}) along with (\ref{eom1}).
We see from the Lagrangian, that the  momentum conjugate to $a$ is
\begin{equation}
p = {\partial L\over{\partial \dot a}} = M_p \sinh^{-1}\Bigg[{\dot
a\over{\sqrt {h(a)}}}\Bigg].
\label{mom}
\end{equation}
It is then straightforward to construct the hamiltonian,
\begin{equation}
H = p \dot a - L = M_p \Bigg[{\sqrt{h(a)}} \cosh \Bigg({ p \over M_p}\Bigg) -
{T
\over{D-1}}a\Bigg].
\label{ham}
\end{equation}
We will, in section 4, use this hamiltonian with suitable ordering to 
analyse a possible quantum scenario for the brane universe. However, 
before doing so, we will study the classical equation of motion of $a$
in the next section.

\noindent{\bf Bulk with dS asymptotics}

Brane evolution  in de Sitter black hole background can simply be obtained 
from (\ref{valueh}) by replacing $l^2 \rightarrow -l^2$ and setting $k 
=1$. In this case, $h(a) =0$ can have three positive roots. The largest 
one corresponds to the cosmological horizon and the rest two correspond to 
the inner and outer horizons of the black hole \cite{LR}. The brane
equations remain same as in equations (\ref{eom1}) and (\ref{eqm}) with 
the hamiltonian as in (\ref{ham}).

\section{Classical cosmological evolution of the brane}~\label{crit}
The classical cosmological evolution of the brane is determined from the
solutions of (\ref{eom1}). Substituting the form of $h(a)$ from
(\ref{valueh}) in (\ref{eom1}), we get  
\be 
\Bigg(\f{\dot a}{a}\Bigg)^2=-\f{k}{a^2}+\Lambda_D+\f{\omega_D
  M}{a^D}-\f{(D-1)\omega^2_D Q^2}{8(D-2)a^{{2D-2}}},
\label{Hubble}
\ee
where $\Lambda_D=\f{T^2}{(D-1)^2}-\f{1}{l^2}$ is the effective
cosmological constant on the brane. Note that for  AdS bulk, the effective 
cosmological constant can be either zero, negative or positive. However,
for dS bulk, $\Lambda_D$ is strictly positive.

We  can now find solutions of (\ref{Hubble}) in
different situations, namely, for different values to $k$,
for different choices of $\Lambda_D$ (positive, negative and zero). We 
thus have three distinct cases:
$(1)~ Q=0, M=0,~ (2)~ Q=0, M >0$ and $(3)~ Q >0, 
M >0$ and for each of these three cases one have the following 9 situations
\bea
\label{27_situations}
& &\Lambda_D=0,\quad k=0,\pm 1\nonumber \\
& &\Lambda_D >0,\quad k=0, \pm 1\nonumber \\
& &\Lambda_D <0, \quad k=0, \pm 1.\nonumber \\
\eea
We will now discuss all these cases in the following.
%

\noindent ${\bf (1)~ {M = Q =0} }$

\noindent For $M = Q =0$, the brane universe moves in AdS/dS bulk. Solving
(\ref{Hubble}), we get the time dependence of the scale factor of the
brane as
\begin{eqnarray}
a &=& {1\over \sqrt{\Lambda_D}}e^{{\sqrt{\Lambda_D}} \tau} ~{\rm for} ~ k 
=0,
\nonumber \\
  &=& {1\over \sqrt \Lambda_D} {\rm cosh} \Big({\sqrt 
\Lambda_D}\tau\Big)~{\rm
for} ~ k =+1,
\nonumber\\
  &=& {1\over{\sqrt\Lambda_D}}{\rm sinh} \Big({\sqrt\Lambda_D} \tau\Big) 
~{\rm for}
~ k = -1.\nonumber\\
\label{solone}
\end{eqnarray}
The above solutions are valid only for $\Lambda_D >0$.
For $k=0$, we get a de-Sitter like
expansion.
For $ k =1$, the universe initially is infinitely large,
then it shrinks
at finite time to a finite size controlled by $\sqrt \Lambda_D$ and,
subsequently, expands and reaches infinite size at late time $\tau$. 
For $k = -1$, the radius starts out from zero size and then grows to
infinite size with time. 

\noindent For the case of $\Lambda_D < 0$, there are no physically 
relevant solutions except for $k = -1$, where $a$ behaves as 
\begin{equation}
\label{k=-1_-velambda}
a = {1\over {\sqrt {|\Lambda_D|}}}{\rm sin} ({\sqrt 
{|\Lambda_D|}}\tau).
\end{equation}

\noindent ${\bf (2)~ M > 0, Q =0}$

\noindent In this case, the brane moves in Schwarzschild black hole with
AdS/dS
asymptotics. The effective brane cosmological constant $\Lambda_D$
can be either zero or non-zero. We discuss both these cases for the $D+1$
bulk in the following. Though in many cases exact solutions can be
obtained, we resort to numerical analysis where exact results
are either hard to obtain or less illuminating.

\noindent ${\bf (i) \Lambda_D = 0}$

\noindent  It turns out that, when the effective brane cosmological
constant is zero, the exact solutions are most
easily obtained in terms of conformal time $\eta$ defined as 
$a(\eta) d\eta = d\tau$. The solutions of (\ref{Hubble}) then gives, 
\begin{eqnarray}
\label{q=0_nonzero_m_zero_lambda}
a(\eta) &=& \Big[{ {D-2\over{2}}{\sqrt{M\omega_D}}} 
\eta \Big]^{2\over{D-2}} ~{\rm for}~k
= 0, \nonumber \\
  &=& (M\omega_D)^{1\over{D-2}}{\rm sin}[{D-2\over{2}}\eta]
^{2\over{D-2}} ~{\rm for} ~k = +1,\nonumber\\
  &=& (M\omega_D)^{1\over{D-2}}{\rm sinh}[{D-2\over{2}}\eta]
^{2\over{D-2}} ~{\rm for} ~k = -1.\nonumber\\
\label{soltwo}
\end{eqnarray}
As can be seen from above, for $k=0$, the brane starts out from
singularity and expands to infinite size at late time. However, the rate
of expansion and acceleration depend crucially on the dimension of the 
brane. For $D=4$, it expands at a constant rate and for $D >4$, the
expansion rate decrease with time $\eta$ producing a
deflationary scenario. In terms of cosmic time $\tau$, the brane 
expands as radiation dominated universe $a \sim \tau^{2\over D}$. For $k =
1$, we get cyclic universe with maximum size of the brane is $a_{max} 
= (M \omega_D)^{1\over{D-2}}$ where $M$ is the  mass of the bulk black 
hole. For $k =-1$, on the other hand, we have an open universe as
expected.

\noindent ${\bf (ii) \Lambda_D \ne 0}$\\

\noindent When the brane has effective cosmological constant (negative or
positive), the dynamics of the universe can be analysed in a quite simple
manner. We  discuss first the case of positive cosmological constant
and then the case with $\Lambda_D <0$. 

\noindent ${\it (a) \Lambda_D > 0}$\\

\noindent In this case, (\ref{Hubble}) can be solved exactly in arbitrary
$D$ for $k =0$ with the result,
\begin{eqnarray}
a(\tau) &=& \Big({{M\omega_D}\over{\Lambda_D}}\Big)^{1\over{D}}{\rm sinh}[{{D\sqrt
{\Lambda_{D}}}\over{2}}\tau]^{2\over{D}}
 ~{\rm for} ~k = 0.\nonumber\\
\label{kzero}
\end{eqnarray}
Therefore, the brane here emerges from the bulk singularity and then
expands for ever. In particular, at late time, we have de-Sitter like 
inflation as can be seen from expanding (\ref{kzero}) for large $\tau$.

\noindent However, for $k = \pm 1$, for only $D= 4$, exact analytical
result can be found and is discussed for $k =1$ in some detail in 
\cite{PS, Med}. For
other
dimensions we could get the behavior of
the scale factor through numerical analysis. For $D=4$, we get
\begin{eqnarray}
a (\tau) &=& \Big[{e^{2 {\sqrt{\Lambda_4}}\tau} + k\over
{2\Lambda_4}}\Big]^{1\over 2} ~{\rm for}~ 4 \Lambda_4 \omega_4 M =1,
\nonumber \\ 
 &=& {1\over {\sqrt{2 \Lambda_4}}}\Big[{\sqrt{1 - 4 \omega_4 M \Lambda_4}}
~{\rm cosh} (2 {\sqrt \Lambda_4} \tau) + k\Big]^{1\over 2}
~{\rm for}~ 4 \omega_4 M\Lambda_4 < 1,\nonumber\\
&=& {1\over {\sqrt{2 \Lambda_4}}}\Big[{\sqrt{4 \omega_4 M \Lambda_4-1}}
~{\rm sinh} (2 {\sqrt \Lambda_4} \tau) + k\Big]^{1\over 2}
~{\rm for}~ 4 \omega_4 M\Lambda_4 > 1.\nonumber\\
\label{solthree_a}
\end{eqnarray}
Thus, for the case $4 \Lambda_4 \omega_4 M =1$, we get for de-Sitter
expansion for all values of $k$ for large $\tau$. However, for small
$\tau$, expansion is sensitive to $k$. 
For $k =1$ the brane starts from a {\it
finite} size $1\over{{\sqrt 2\Lambda_4}}$, while for $k =-1$, brane again 
emerges
from bulk singularity. As can be seen from (\ref{solthree_a}), 
when $4 \Lambda_4 \omega_4 M < 1$, the universe radius is {\it bounded}
from
{\it below} by $({\sqrt{1 - 4 \Lambda_4 \omega_4 M}} + k)^{1\over
2}/{\sqrt{2 \Lambda_4}}$ for $ k = 1$. 

\noindent As mentioned earlier, for $D > 4$, analytical solutions in 
closed forms are
hard to obtain. However, below are the plots
of $a$ with time for different $k$.\\

\begin{figure}[ht]
\epsfxsize=10cm
\centerline{\epsfbox{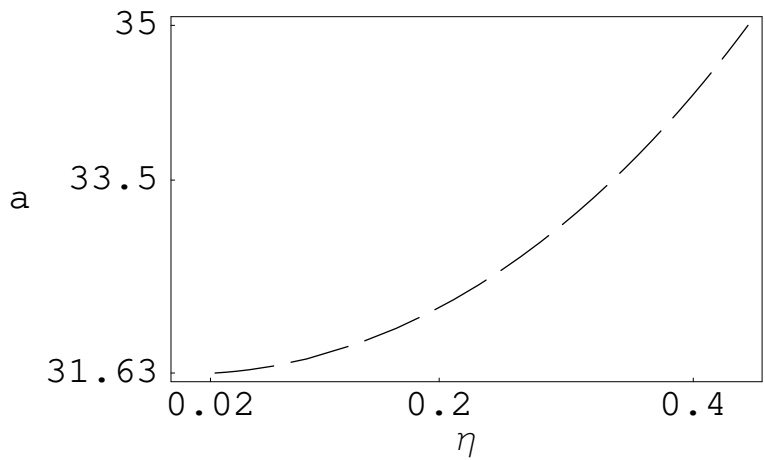}}
\caption{The universe for $D =6$ and $k = 1, ~\omega_6 M =100, ~\Lambda_6 
= .001$. 
It begins at a finite size and expands for ever.
Here and in other figures, all the quantities are measured in 
appropriate powers of length.}
\end{figure}

\begin{figure}[ht]
\epsfxsize=10cm
\centerline{\epsfbox{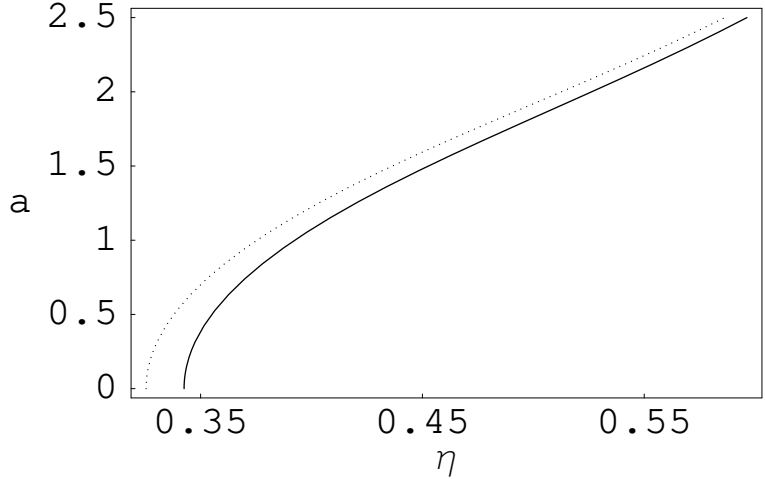}}
\caption{The universe for $D =6$ and  $\omega_6 M =100,
~\Lambda_6 = 1.$ Here $k =-1$ corresponds to the solid line while
dotted
line is for $k =0$. In both cases, the universe begins from 
singularity and expands for ever. }
\end{figure}

\newpage
\noindent ${\it (b) \Lambda_D < 0}$ \\

\noindent The case when effective cosmological constant is negative on
the brane, (\ref{Hubble}) can be exactly integrated for arbitrary $D$ for 
flat universe with the result
\begin{eqnarray}
a(\tau) = \Big({{M\omega_D}\over{{\Lambda}_D}}\Big)^{1\over{D}}
{\rm sin}[{{D\sqrt{\Lambda_D}}\over{2}}\tau]^{2\over{D}} 
~{\rm for} ~k = 0. \nonumber \\
\label{lamneg}
\end{eqnarray}
We therefore have cyclic universe with minimum radius being zero while the
maximum is  determined by the ratio of black hole mass and cosmological
constant.

\noindent As before, for $k = \pm 1$, exact solution can be found easily 
for
$D = 4$. Those are given by
\begin{equation}
 a(\tau)  = {1\over\sqrt{2\Lambda_4}}\Big[
{\sqrt{4M\omega_4{\Lambda_4}+1}
~{\rm sin}(2\sqrt{\Lambda_4}\tau) - k}\Big]^{1\over2} 
~{\rm for} ~k = \pm 1.
\label{solthree_b}
\end{equation}

\noindent For $D >4$, the behavior of $a$ can be found by numerical means
and is shown in the following figure.
\begin{figure}[ht]
\epsfxsize=10cm
\centerline{\epsfbox{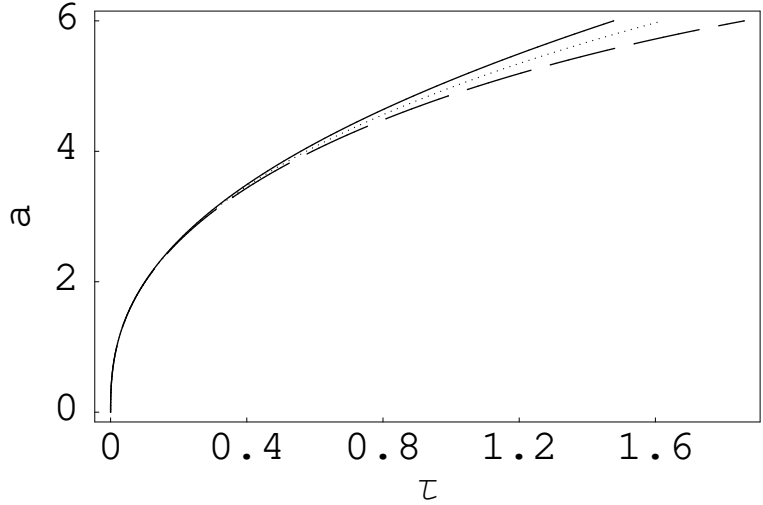}}
\caption{Evolution of the brane in $D=5$ for $\omega_5 M = 500,~\Lambda_5 
=
.01$. The dotted, dashed and solid lines correspond to $k = 0, 1, -1$
respectively.} 
\end{figure}

\noindent ${\bf (3) M > 0,~ Q > 0}$ \\

\noindent In this subsection, we turn our attention to the case where the 
brane is moving in charged AdS background parametrized by mass $M$ and
charge $Q$. In this case, as will be discussed below, we get a host of
brane cosmologies which are completely {\it non-singular} during
the brane evolution.
\newpage
\noindent ${\bf (i) \Lambda_D = 0}$

\noindent We start here with the case, where the effective cosmological
constant $\Lambda_D$ is zero. In this case, (\ref{Hubble}) can be exactly
solved for any $D$ with the result
\begin{eqnarray}
a(\eta) &=& \Big[{1\over M\omega_D}\Big({(D-1)Q^2\omega_D^2\over{8(D-2)}}+
{{M^2\omega_D^2\over 4}(D-2)^2}\eta^2\Big)\Big]^
{1\over D-2} ~{\rm for} ~k = 0, \nonumber \\
  &=& \Big[{M\omega_D\over 2} \Big(1+\sqrt{1-{D-1\over 2(D-2)}
{Q^2\over M^2}}{\rm sin}((D-2)\eta)\Big)\Big]^{1\over D-2}
~{\rm for} ~k = +1, \nonumber \\
  &=& \Big[{M\omega_D\over 2} \Big(-1+\sqrt{1+{D-1\over 2(D-2)}
{Q^2\over M^2}}{\rm cosh}((D-2)\eta)\Big)\Big]^{1\over D-2}
~{\rm fo}r~k = -1. \nonumber \\
\label{solfour}
\end{eqnarray}
It is important to note that $k=1$ solution only makes sense when the
charge 
satisfies the inequality
\begin{equation}
Q \le {\sqrt 2} M {\sqrt {D -2\over{D-1}}}.
\end{equation}
For $D=4$, these configurations are studied in some detail in 
\cite{MP}. 
As can be seen from the figure 4, all three solutions are {\it 
non-singular}. For $k=0, -1$, universe starts from nonzero value of 
$a$  and expands forever. 
Thus for flat brane
\begin{equation}
\Big[{1\over M\omega_D}{(D-1)Q^2\omega_D^2\over{8(D-2)}}\Big]^{1\over {D -2}} \le a 
\le
\infty,
\end{equation}
and for $k = -1$
\begin{equation}
\Big[{M
\omega_D\over 2} \Big(-1 + {\sqrt {1 + {(D-1)Q^2 \over{2 (D-2)
M^2}}}}\Big) \Big]^{1\over{D-2}} \le a \le \infty.
\label{k-rad}
\end{equation}
On the other hand, we have bouncing 
non-singular universe for $k =1$  where the universe 
oscillates between two non-zero values of $a$ as
\begin{equation}
\Big[{M 
\omega_D\over 2} \Big(1 - {\sqrt {1 - {(D-1)Q^2 \over{2 (D-2) 
M^2}}}}\Big) \Big]^{1\over{D-2}} \le a \le
\Big[{M \omega_D\over 2} \Big(1 + {\sqrt {1 - {(D-1)Q^2 \over{2 (D-2)
M^2}}}}\Big) \Big]^{1\over{D-2}}.
\end{equation}
Notice that as we take $Q$ to zero, the minimum radius collapses to 
a singularity. So, in a sense, the charge parameter $Q$ acts as a 
regulator.
This behavior is also intuitively expected. As in this case, the brane 
moves in electrically charged background, the flux lines have to end on 
the brane. These fluxes, in turn, do not allow the brane to shrink to zero 
size. A typical behavior of the universe for $ k = 0, \pm 1$ is shown in 
the following figure. \\\\\  
\begin{figure}[ht]
\epsfxsize=10cm
\centerline{\epsfbox{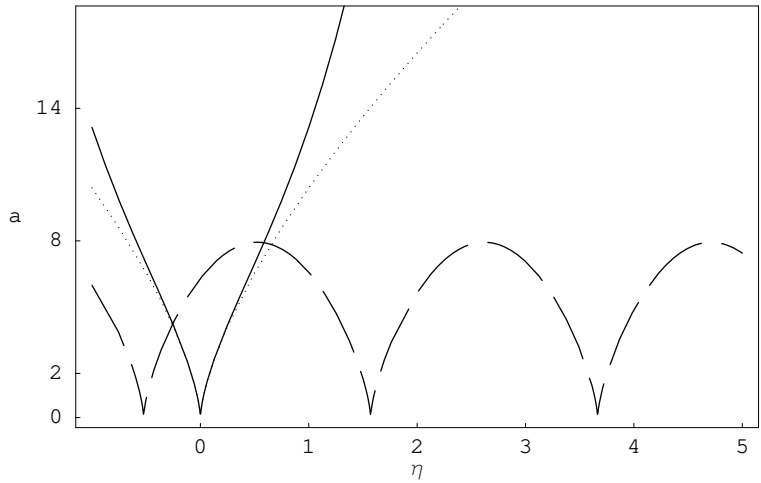}}
\caption{The universe for $k = -1, 0, 1$ are shown in solid, dotted and
dashed lines respectively for $D = 5, ~\omega_5 M = 500, ~\omega_5 Q=1$ 
and $\Lambda_5 = 0$. 
As can be seen, in 
all these cases, big bang like singularities are absent.} 
\end{figure} \\
\newpage
We further would like to emphasize that for $k =-1$ and $k =1$, the 
minimum size of the universe can be made sufficiently large by taking bulk 
mass and charge to large values. Thus, one would expect to have quantum 
gravity corrections to be small even when the brane is at its 
minimum size. In this sense, the universe avoids the usual quantum gravity 
era in the past.

\noindent ${\bf (ii) \Lambda_D \ne 0}$

It turns out that in this case, it is hard to integrate (\ref{Hubble}) 
to get a closed form expression for $a$ as a function of $\tau$. However, 
the equation can easily be integrated numerically. In what follows, we 
discuss the solutions for $\Lambda_D >0$ as well as $\Lambda_D <0$ for 
$D =4$.

\noindent ${\it (a) \Lambda_D > 0}$

As can easily be checked from (\ref{Hubble}) that for fixed $\Lambda_D$, 
for certain range of parameters $(M,Q)$, $\dot a$ becomes zero at a 
positive value of $a$ at some finite time $\tau$ for all $k$. The universe 
then begins at that radius. At late time, brane expands exponentially 
independent of the values of $k$. The expansion rate of course depends on 
the effective cosmological constant $\Lambda_D$. The complete evolution of 
the brane is shown in the figure below for certain choice of parameters.\\

\begin{figure}[ht] 
\epsfxsize=10cm
\centerline{\epsfbox{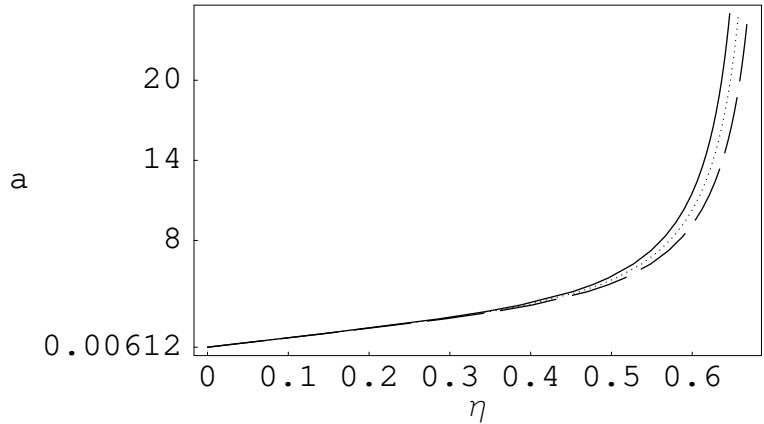}}
\caption{This shows typical behavior of the universe for $\Lambda_4 =1$.
Here again $k =-1, 0, 1$ are shown in solid, dotted and dashed lines for
$D=4,~\omega_4 M = 50, ~\omega_4 Q =0.1$. Universe begins at a finite size 
and expands at
late time.}
\end{figure}
\newpage
\noindent ${\it (b) \Lambda_D < 0}$
 
In the case where the effective cosmological constant is negative, for 
fixed $\Lambda_D$, for certain choice of parameters $(M,Q)$, there are two 
real positive roots of the equation $\dot a = 0$ for all $k$. The universe 
then begins then at finite $a_{min}$ and ends at another value of 
$a_{max}$. Below $a_{min}$ and above $a_{max}$, the brane is unstable. 
The complete evolution of the brane for certain choice of parameters is
shown in the figure below.\\\\
\begin{figure}[ht]
\epsfxsize=10cm
\centerline{\epsfbox{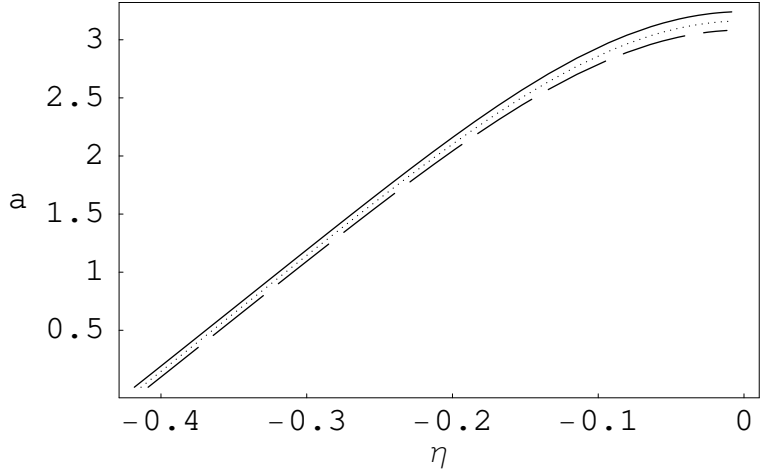}}
\caption{This shows typical behavior of the universe for $\Lambda_4 = 
-1$. 
$k =-1,0,1$ are shown in solid, dotted and dashed lines for $D 
=4,~\omega_4 M
=100,~\omega_4 Q = 0.1$. Universe begins at non-zero size and ends at 
finite size. 
Instabilities arise below the minimum and above the maximum radii.}
\end{figure}

As we have now seen, in some instances, universe begins at the
singularity of the bulk. This in turn imply big bang like singularity on
the brane. In the next section, we study these singular universes from the
perspective of ``quantum cosmology".


\section{Quantum cosmologies arising from Wheeler-De Witt
Equation}~\label{radio}
In this section, we turn our attention to a quantum version of the 
cosmological models of previous section. Due to lack of an existing 
formulation of quantum gravity, our task becomes really difficult. We, 
therefore,  need to make drastic simplifications. The approximation that 
is most 
commonly used is the mini superspace truncation of the Wheeler-De Witt 
formalism. In what follows, we will make such approximation in order to 
discuss a suggestive quantum version of our previous models. 
Even with in the mini-superspace approximation, quantum cosmology suffers
from many ambiguities. The operator ordering ambiguity in the hamiltonian
operator, choice of boundary condition on the wave function are among many
others. We may say that we have nothing to add to improve our
understanding on those issues. We rather work with a particular operator
ordering and chose a particular boundary condition on the wave function
which seems natural to us. As a result, our result in this section would
at most be speculative in nature. We will 
closely follow the method developed in \cite{Vis}. Creation of 
brane-universe in AdS space is discussed in \cite{ANO}.

First, the Wheeler-De Witt hamiltonian  can be obtained from (\ref{ham}) by
substituting $p\rightarrow -i\f{\partial}{\partial a}$ and is  
\be 
{\hat H}=M_p\Bigg[
h^{\f{1}{4}}~~\cos\Big({\f{1}{M_p}\f{\partial}{\partial
    a}}\Big)~~h^{\f{1}{4}}-\f{T}{(D-1)}a\Bigg], 
\label{Hord}
\ee
and the evolution of the brane universe  is then governed by
\be 
{\hat H}\psi(a)=0.
\label{caph}
\ee
Here, $\psi (a)$ is the wave function of the brane universe 
while $\hat H$ is a 
quantum hamiltonian of the brane.
In writing down (\ref{Hord}), one faces the ordering 
ambiguity. Hermiticity of the hamiltonian is used as guideline to set this 
operator ordering. We, however, may note that the ordering here is not 
unique and one can, in principle, work with a different ordering prescription. 
This, in turn, will change the nature of the quantum behavior.

It is now straightforward to solve (\ref{caph}) particularly for $T=0$. 
The  solution is 
\be
\psi_{mn}(a)=C_{mn}(\phi_m-\phi_n),
\ee
with 
\be 
\phi_j=h^{-\f{1}{4}} ~~{\rm exp}\Bigg[-\Big(j+\f{1}{2}\Big){\pi a 
M_p}\Bigg].
\ee
$C_{mn}$ here are normalisation constants which can be explicitly 
evaluated. 
However, for our purpose, we would not require their explicit forms. In 
the above expression of the wave functions,
$m, n$ are the integer valued quantum numbers describing the 
state of the brane.  
Negatives values of $m$, $n$ do not give
normalizable wave function so we discard them. Same is the case for
$m=n$. It is now quite easy to see that for both $M=0 ~{\rm and} ~Q=0$, 
and 
$M\neq 0, Q\neq 0 $, the
solution, as written above, satisfies both the boundary condition
(given later) and Wheeler-De Witt  equation. However, for $M \neq 0, Q\neq 
0$, we  have another solution, which also satisfies both Wheeler-De Witt 
equation and
the boundary condition,  namely,
\be 
\psi_j(a)=C_{j}\phi_j,
\ee
where $\phi_j$ is as written above and $C_j$ is again a normalisation 
constant.

Note that our wave function is normalised as $\int^{\infty}_{0}|\psi|^2 da
=1$. 
Furthermore, we have chosen De-Witt boundary (see for example 
\cite{tipler}) condition on the wave
function.
\begin{equation}
\psi(a=0) = 0.
\label{bcond}
\end{equation}
This simply says that the probability of  occurrence of the singularity 
at $a=0$ is zero. We will continue to use this boundary condition 
on the wave function in our following analysis. Existence of a normalised
wave function with this boundary condition is then
suggestive to the fact that the universe may not have the initial
singularity in the quantum version.
Let us also note that the two terms in $\psi_{mn}$
individually satisfy the differential equation ${\hat H}
 \psi =0$, but do not individually satisfy the boundary
condition for $M=0$ and $Q=0$.
With the wave functions at hand, one can calculate the mean value of the
``radius'' of the brane universe, that is
\begin{equation}
\langle a\rangle=\frac{\int a |\psi_{mn}|^2 da}{\int|\psi_{mn}|^2 da}.
\label{meanvalue}
\end{equation}
By dimensional analysis this number can be expected to be the order of 
$\frac{1}{M_p}$. We can find the
solution for $T=0$, that is for
$\Lambda_D=-\frac{1}{l^2}$. Using the form $h(a)$ as  
\be 
h(a)=k-\Lambda_D a^2-\frac{\omega_D M}{a^{D-2}}+\frac{(D-1)\omega^2_D
  Q^2}{8 (D-2) a^{2D-4}}.
\ee
Hence, the solution $\phi_j$ is
\be 
\label{solution_phi}
\phi_j=\frac{exp[-(j+\frac{1}{2})\pi a M_p]}{(k-\Lambda_D a^2-\frac{\omega_D M}{a^{D-2}}+\frac{(D-1)\omega^2_D
  Q^2}{8 (D-2) a^{2D-4}})^{\frac{1}{4}}}.
\label{thirtye}
\ee
It is important to note that the wave function blows up when the 
denominator of (\ref{thirtye}) vanishes for non-zero values of $a$. 
This happens at the horizon of the bulk black hole. The
horizon singularity, in black hole physics,  
can be removed by proper choice of coordinates. Therefore, such 
divergences of the wave function may not be so worrying for us (see
\cite{Vis} for a discussion on this issue). 
%
On the other hand, the real 
singularity of the bulk is at $a =0$. It is quite easy to check that the 
wave function approaches zero at this singularity satisfying the boundary 
condition (\ref{bcond}).  

Let us now proceed to analyse some consequences of the wave function we 
have just found for some specific cases. We first consider the case where 
$M = Q = 0, k = -1$ and $\Lambda_D <0$. The classical behavior of the 
brane was analysed in the previous section. The universe then starts from 
singularity, expands till the maximum size $\frac{1}{\sqrt {|\Lambda_D|}}$
and then shrinks to singularity. From the quantum mechanical perspective, 
we find the wave function to be 
\be 
\psi_{mn}(a)=\frac{C_{mn}}{\bigg(-1+|\Lambda_D |a^2\bigg)^{\frac{1}{4}}}
\Bigg( e^{-(m+\frac{1}{2})\pi a M_p}-e^{-(n+\frac{1}{2})\pi a M_p}\Bigg),
\ee
Consequently, we can evaluate the expectation value of the universe radius. 
This turns out to be
\be 
\langle a\rangle= \frac{\int^{\infty}_{0} da a |\psi_{mn}|^2}
{\int^{\infty}_{0} da  |\psi_{mn}|^2}.
\ee
Further, expanding the denominator of $\psi_{mn}$ and keeping terms to  
linear in 
$\Lambda_D$ 
with $m=1, n=0$ gives 
\be 
\langle a\rangle=\frac{1}{\pi M_p}
\Bigg(\frac{11}{6}+\frac{\Lambda_D}{(\pi M_p)^2}\frac{395}{108}\Bigg)+
{\cal O}({\Lambda_D}^2).
\ee  
Similarly, 
for  $M\neq 0,Q=0, k=1$ and $\Lambda_D=0$ (with  $T=0, l\rightarrow\infty$), 
as written in the second line of eq. (\ref{q=0_nonzero_m_zero_lambda}) and is 
\be 
\psi_{mn}(a)=\frac{C_{mn}}{\bigg(1-\frac{\omega_D M}{a^{D-2}}\bigg)^{\frac{1}{4}}}\Bigg( e^{-(m+\frac{1}{2})\pi a M_p}-e^{-(n+\frac{1}{2})\pi a M_p}\Bigg),
\ee 
the average value of $a$ for small $\omega_D M$ with $m=1$ and $n=0$ is
\begin{eqnarray} 
\langle a\rangle &=&\frac{11}{6 \pi M_p}+\frac{3\omega_D M}{2}
\frac{\Gamma(4-D)}{(\pi M_p)^{3-D}}\bigg(\frac{1}{3^{4-D}}+1-
\frac{1}{2^{2-D}}\bigg) \nonumber \\
&-&\frac{11\omega_D M}{4}\frac{\Gamma(3-D)}{(\pi M_p)^{3-D}}\bigg(
\frac{1}{3^{3-D}}+1-\frac{1}{2^{2-D}}\bigg)+{\cal O}(\omega_D M)^2.
\end{eqnarray}
We thus notice that in both the above cases, the universe radius gets an 
average value which, in leading order, is proportional to $1/M_p$ and is 
{\it independent} of the other dimensionful constants that appear in 
$\psi_{mn}$. One can further evaluate $<a>$ for small but non-zero 
$\Lambda$. We have explicitly checked that the leading order term is again 
of the order $1/M_p$ and free of other dimensionful parameters.
This may indicate that the universe, in stead of shrinking to zero size, may 
stabilize at the scale of $1/M_p$ due to quantum effects. 
In the next section, we would study the solution for nonzero $T$ using WKB 
approximation. 

\section{WKB Approximation}

It turns out to be  too difficult to find the solution of the Wheeler-De 
Witt equation for $T\neq 0$. 
However, by employing WKB technique, it is possible to get a suggestive 
picture of the quantum universe for branes with non-zero tension.
 
Let us recapitulate the recipe to find the wave function of a system 
using WKB 
approximation. 
First, we have to solve the the equation 
${\hat H({\hat p},{\hat a})}\psi=E\psi$, with $H(p,a)$ as 
the Hamiltonian associated to  
classical system.  Setting this classical Hamiltonian, $H(p,a)=E$, gives us 
a relation between $p=f(E,a)$. Using this approximation, we can find 
solution 
to the Hamiltonian equation for two regimes: for $E > V$ and $E < V$,  the 
 former corresponds to the classically allowed region and the latter to 
classically forbidden region. The solutions for these cases are
\bea 
\psi_A(a)&=&\frac{1}{\sqrt{(\frac{\partial H}{\partial p})}}
\times e^{\pm i\int^a p dx}\quad{\rm 
~~for ~~classically ~~allowed ~~region,} \nonumber \\
\psi_F(a)&=&\frac{1}{\sqrt{|(\frac{\partial H}{\partial p})|}}
\times e^{- \int^a |p| dx}\quad{\rm 
~~for ~~classically ~~forbidden ~~region.} 
\eea

Following the prescription as written above, it is quite easy to find the 
momentum $p$. It  is given by 
\bea 
p&=&M_p \cosh^{-1}\bigg[ \frac{Ta}{(D-1){\sqrt {h(a)}}}\bigg]\nonumber \\
&=& M_p \bigg(\cosh^{-1}\bigg[ \frac{Ta}{(D-1){\sqrt {h(a)}}}\bigg]+2 \pi in\bigg ), 
\eea
where $n$ is an integer. The conjugate momentum, as written above, has to 
satisfy the Bohr-Summerfield quantization condition
\be 
\oint p da= (m+\delta),
\ee  
where $m$ is an integer and associated to ``radial'' quantum number, and 
$\delta$ is typically a fraction. We will not require the explicit
value of $\delta$.
The wave function for the classically 
allowed region is then
\bea
\label{solution_allowed}
\psi_A(a)&=&\frac{1}{{\sqrt{\dot a}}} \times e^{ i\int^a p dx} \nonumber \\
&=&\frac{e^{-2\pi naM_p}}{\bigg(-k+a^2\Lambda_D+\frac{\omega_D M}{a^{(D-2)}}-\frac{(D-1)\omega^2_D Q^2}{8(D-2)a^{{2D-4}}}\bigg)^\frac{1}{4}}\times e^{i\Theta(a)},\nonumber \\
\eea
where $\Theta(a)$ is 
\be 
\label{theta}
\Theta(a)=M_p\int^a \cosh^{-1}\bigg[ \frac{Ta}{(D-1){\sqrt {h(a)}}}\bigg].
\ee
It is easy to see from (\ref{solution_allowed}) that the solution in 
the classically allowed region is oscillatory in nature and simultaneously 
exponentially damped for $n > 0$. The solution in the classically forbidden 
region is 
\be 
\psi_F(a)=\frac{1}{\sqrt{|\dot a|}} e^{-2\pi n aM_p}e^{-|{\Theta(a)} |},
\ee
where $\Theta(a)$ is given in eq. (\ref{theta}). 
\section{Discussion}~\label{pert}
In this paper, we have initiated a systematic study of possible 
cosmological scenarios on a brane where the brane acts as a dynamical 
boundary of various asymptotically AdS or dS spaces. Our emphasize, thought out, 
was to isolate the cases where big bang singularity was absent on the 
brane. This happens, in particular, when the bulk represents $D +1$ 
dimensional charged AdS or dS black hole. Early time singularities for $k 
= 0, \pm 1$ are found to be absent on the brane. This was found by 
explicitly solving the FRW equation when the effective cosmological 
constant on the brane was fine tuned to zero. Furthermore, by studying FRW 
equation through numerical means, we found that the non-singular feature 
continues even when the effective cosmological constant is non-zero. 
We have also found that, by taking mass and the charge of the bulk black 
hole to large values 
(but keeping their ratio fixed), the universe can avoid quantum gravity 
era at early time for the models with $k = \pm 1$. In section 4 and 5, we 
analysed the brane universe from  the quantum mechanical perspective. This 
was done by employing Wheeler-De Witt equation in the mini superspace 
formalism and also via WKB methods. 
Within these approximation schemes, we studied the wave function of the 
brane-universe with a special attention to the scenarios where classical 
big bang singularities are present. We then found that the minimum size of 
the 
brane is of the order of Planck length. 

As it is clear that many interesting issues of cosmological importance 
were not considered in this paper. No matter fields (except those that 
were induced from bulk) were introduced on the brane. These fields 
would certainly modify the scenarios that we have discussed so far. 
It would be nice to see, however, if the scenario of bounce followed 
by radiation still persists when such matter fields are included.
Furthermore, fluctuation of such matter fields also remain an interesting 
arena to explore. We hope to address these issues in the future.

\vskip 0.5cm

\noindent{\large\bf Acknowledgment:} We wish to thank Rajarshi Ray for 
discussions. SM would like to thank Stefan Theisen for hospitality at AEI, 
Potsdam during the initial stages of the work.


\newpage


\begin{thebibliography}{99}

\bibitem{HW} P. Horava and E. Witten, Nucl. Phys. {\bf B460} (1996) 460,
hep-th/9510209.


\bibitem{JM} J. Maldacena, Adv. Theor. Math. Phys. {\bf 2} (1998), 231,
hep-th/9711200.

\bibitem{ADD} N. Akrani-Hamed, S. Dimopoulos and G. Dvali, Phys. Lett.
{\bf B436} (1998) 263, hep-ph/9804398.

\bibitem{AADD}
I.~Antoniadis, N.~Arkani-Hamed, S.~Dimopoulos and G.~R.~Dvali, Phys.\ 
Lett. {\bf B436} (1998) 257, hep-ph/9804398.

\bibitem{RSo} L. Randall and R. Sundrum, Phys. Rev. Lett. {\bf 83} (1999)
3370, hep-ph/9905221.

\bibitem{RSt}  L. Randall and R. Sundrum, Phys. Rev. Lett. {\bf 83} (1999)
4690, hep-th/9906064.
 
\bibitem{gata}
J. Garriga and T. Tanaka, Phys. Rev. Lett. {\bf 84} (2000) 2778, 
het-th/9911055.

\bibitem{TN}T. Nihei, Phys. Lett. {\bf B465} (1999) 81, hep-ph/9905487.

\bibitem{CG} C. Csaki, M. Graesser, C. Kolda and J. Terning,  Phys.
Lett. {\bf B426} (1999) 34, hep-ph/9906513.

\bibitem{PK} P. Kraus, JHEP {\bf 9912} (1999) 011, hep-th/9910149.

\bibitem{BD} P. Binetruy, C. Deffayet, U. Ellwanger and D. Langlois,
Phys. Lett. {\bf B477} (2000) 285, hep-th/9910219.

\bibitem{DI} D. Ida, JHEP {\bf 0009} (2000) 014, gr-qc/9912002.

\bibitem{MV1} C. Bercelo and M. Visser, Phys. Lett. {\bf B482} (2000) 183,
hep-th/0004056.

\bibitem{ANO}
L.~Anchordoqui, C.~Nunez and K.~Olsen, JHEP {\bf 0010} (2000) 050, 
hep-th/0007064.

\bibitem{no}
S.~Nojiri and S.D.~ Odintsov, Phys. Lett. {\bf B 484} (2000) 119, hep-th/0004097.

\bibitem{noo}
S.~Nojiri, O.~Obregon and S.D.~ Odintsov, Phys. Rev. {\bf D 62} (2000) 104003, hep-th/0005127.


\bibitem{BCG} P. Bowcock, C. Charmouis and R. Gregory, Class. Quant. Grav. 
{\bf 17} (2000) 4745, hep-th/0007177.

\bibitem{CEG} C. Csaki, J. Erlich and C. Grojean, Nucl. Phys. {\bf B604}
(2001) 312, hep-th/0012143.

\bibitem{no1}
S.~Nojiri and S.D.~ Odintsov, Class. Quan. Grav. {\bf 18} (2001) 5227, hep-th/0103078.

\bibitem{MY}
Y.~S.~Myung, hep-th/0103241.

\bibitem{CMO} R.G. Cai, Y.S. Myung and N. Ohta, Class. Quant. Grav. {\bf 
18} (2001) 5429, hep-th/0105070.

\bibitem{ap}
A.~Padilla, Phys. Lett. {\bf  B528} (2002) 274, hep-th/0111247.

\bibitem{CO}
D.~H.~Coule,
Class.\ Quant.\ Grav.\  {\bf 18} (2001) 4265.

\bibitem{vs}
Y.~Shtanov and V.~Sahni, gr-qc/0208047.


\bibitem{ap1}
A.~Padilla, hep-th/0210217.

\bibitem{GP}
J.~P.~Gregory and A.~Padilla,
Class.\ Quant.\ Grav.\  {\bf 19} (2002) 4071,
hep-th/0204218.

\bibitem{NOO}
S.~Nojiri, S.~D.~Odintsov and S.~Ogushi, hep-th/0205187.


\bibitem{cm}
R.G.~Cai and Y.S.~Myung, hep-th/0210272.

\bibitem{Gub} S. S. Gubser, Phys. Rev. {\bf D63} (2001), hep-th/9912001.

\bibitem{EV} E. Verlinde, hep-th/0008140.

\bibitem{SV} I. Savonije and E. Verlinde, Phys. Lett. {\bf B507} (2001) 
305, hep-th/0102042.

\bibitem{MP} S. Mukherji and M. Peloso, Phys. Lett. {\bf B547} (2002) 297, 
hep-th/0205180.

\bibitem{BM} A.K. Biswas and S. Mukherji, JHEP {\bf 0103} (2001) 046,
hep-th/0102138.

\bibitem{EM} G. R. Ellis and R. Maartens, gr-qc/0211082.

\bibitem{GH} G.W. Gibbons and  S.W. Hawking, Phys. Rev. D {\bf 15} (1997) 
2752.

\bibitem{BK} V. Balasubramanian and  P. Kraus,  Commun. Math. Phys. {\bf
   208} (1999) 413, hep-th/9902121.

\bibitem{KLS}  P. Kraus, F. Larsen  and R. Siebelink, Nucl.Phys. B
  {\bf 563} (1999) 259-278, hep-th/9906127.

\bibitem{CEJM} A. Chamblin, R. Emparan, C. Johnson and R. Myers, Phys.
Rev. {\bf D60} (1999) 064018, hep-th/9902170.

\bibitem{CS}
R.-G. Cai and K.-S. Soh, Phys. Rev. {\bf D59} (1999) 044013,
gr-qc/9808067.

\bibitem{MV} M. Visser, {\it Lorentzian wormholes: from Einstein to
Hawking}, (American Institute of Physics, Woodbury, 1995).

\bibitem{LR} L. Romans, Nucl. Phys. {\bf B383} (1992) 395,
hep-th/9203018.

\bibitem{PS} A. Petkou and G. Siopsis, JHEP {\bf 0202} (2002) 045,
hep-th/0111085.

\bibitem{Med} A.J. Medved, hep-th/0205037, hep-th/0301010.

\bibitem{Vis} M. Visser, Phys. Rev. {\bf D43} (1991) 402.

\bibitem{tipler} F. Tipler, Phys. Rep. {\bf 137} (1986) 231.

\end{thebibliography}
\end{document}